\documentclass[a4paper,fleqn]{cas-dc}

\usepackage[numbers]{natbib}

\usepackage{amsmath,amsfonts}
\usepackage{algorithmic}
\usepackage{array}
\usepackage{orcidlink}
\usepackage[caption=false,font=normalsize,labelfont=sf,textfont=sf]{subfig}
\usepackage{textcomp}

\usepackage{xurl}
\usepackage{verbatim}
\usepackage{graphicx}
\usepackage{balance}
\usepackage{tikz}
\usetikzlibrary{positioning,arrows.meta}
\usepackage{booktabs}
\usepackage{tabularx}
\usepackage{enumitem}
\usepackage{xcolor}
\def\BibTeX{{\rm B\kern-.05em{\sc i\kern-.025em b}\kern-.08em
    T\kern-.1667em\lower.7ex\hbox{E}\kern-.125emX}}

\usepackage{hyperref}
\hypersetup{
    colorlinks=true,
    linkcolor=blue,
    citecolor=blue,
    urlcolor=black
}

\def\tsc#1{\csdef{#1}{\textsc{\lowercase{#1}}\xspace}}
\tsc{WGM}
\tsc{QE}
\tsc{EP}
\tsc{PMS}
\tsc{BEC}
\tsc{DE}

\makeatletter

\def\ps@firstpage{%
  \let\@oddhead\@empty
  \let\@evenhead\@empty
  \def\@oddfoot{\hfil\thepage\hfil}%
  \let\@evenfoot\@oddfoot
}

\def\ps@cas{%
  \let\@oddhead\@empty
  \let\@evenhead\@empty
  \def\@oddfoot{\hfil\thepage\hfil}%
  \let\@evenfoot\@oddfoot
}
\makeatother

\begin{document}
\let\WriteBookmarks\relax
\def\floatpagepagefraction{0.3}
\def\textpagefraction{.001}
\shorttitle{Bridging the First-Hour Gap in Cyber Incident Response}
\shortauthors{Roshin Sleeba C et~al.}

\title [mode = title]{Bridging the First-Hour Gap: Evaluating AI Reliability and Benchmarking Deficiencies in Cyber Incident Response for Law Enforcement}                  

\author[1]{Roshin Sleeba C}[orcid=0009-0008-3844-2302]
\cormark[1]
\ead{roshinsleebac2002@gmail.com}

\author[1]{Hiran V Nath}[orcid=0000-0001-7881-4694]

\ead{hiranvnath@nitc.ac.in}

\credit{Conceptualization of this study, Methodology, Software, Writing - Original draft preparation}

\affiliation[1]{organization={Department of Computer Science and Engineering, National Institute of Technology Calicut},
                state={Keralam},
                country={India}}

\cortext[cor1]{Corresponding author}
\begin{abstract}
The actions of frontline law enforcement officers in the initial hour of a cyber incident play a vital role in determining the ultimate success of an investigation. The minor mistakes they commit might result in irreversible critical impacts. The integrity of the investigation can be compromised, and the prosecution of cyber criminals can be hindered due to minor mistakes that happen in the initial hour. These are mainly because of the volatile nature of digital artifacts that might lead to procedural errors and evidence attrition. This paper provides a systematic survey of decision-support architectures designed to assist first responders of a cybercrime, categorizing them into playbooks, Large Language Models (LLMs), Retrieval-Augmented Generation (RAG) frameworks, and Agentic AI systems. The survey critically considers the constraints of limited technical proficiency and inconsistent forensic infrastructure in a practical scenario. Our analysis identifies RAG-based systems as a relatively viable intermediate solution due to their natural language adaptability. However, significant risk factors like prompt sensitivity and the potential for confident hallucinations in legal contexts pose a major challenge. Furthermore, we review current benchmarks in cybersecurity and demonstrate that they are not sufficient to capture the specific safety and legal requirements of law enforcement, focusing on the initial hour of the cybercrime. We conclude by arguing for the necessity of a new evaluation benchmark focused on naive query robustness and evidence preservation, so as to ensure that AI-driven guidance aligns with the mandatory demands of judicial proceedings.
\end{abstract}

\makeatletter
\ExplSyntaxOn
\cs_set:Npn \__first_footerline: { }
\ExplSyntaxOff
\makeatother
\maketitle

\section{Introduction}

Cybercrime investigations present a critical operational challenge for law enforcement agencies. In many real-world cases, frontline police officers with limited expertise in cybersecurity are the first responders to cyber incidents. These officers are often required to make crucial decisions within minutes of arriving at an incident scene, frequently without immediate access to digital forensics experts or specialized technical support. The actions taken during this initial phase have a disproportionate impact on the success, admissibility, and long-term viability of the investigation. What happens in the first minutes after initial access can determine whether an incident becomes a breach.

In practice, a large fraction of cybercrime investigations begin with a complaint registered by an employee of an affected organization. Upon arrival, the responding officer must assess the situation based on the artifacts visible on the compromised system with a partial understanding of the surrounding technical environment. Typical victim statements may include observations such as encrypted files with unfamiliar extensions or the sudden appearance of suspicious files across multiple directories. Based on such limited and often non-technical descriptions, the officer must immediately decide whether to isolate the system from the network, what evidence must be photographed or documented, which volatile or semi-volatile artifacts must be preserved most optimally before they are overwritten, and when specialized cybercrime teams should be engaged. These are to be done without disturbing the critical sections of the organization. These decisions are taken under significant time pressure and legal constraints. Improper shutdown, or failure to preserve volatile evidence can compromise the investigation irreversibly. Most existing cyber incident response methodologies assume a level of technical expertise that is unrealistic for non-specialist law enforcement personnel operating in the field.

Several decision-support paradigms have been proposed in the literature, and a few are deployed in practice to support early-stage cybercrime response. This survey systematically examines these approaches. Rather than mere evaluation in abstract technical terms, this survey critically analyzes them through the lens of frontline law enforcement usage by highlighting real-world constraints.

\begin{figure}[pos=t] 
\centering
\includegraphics[width=\columnwidth]{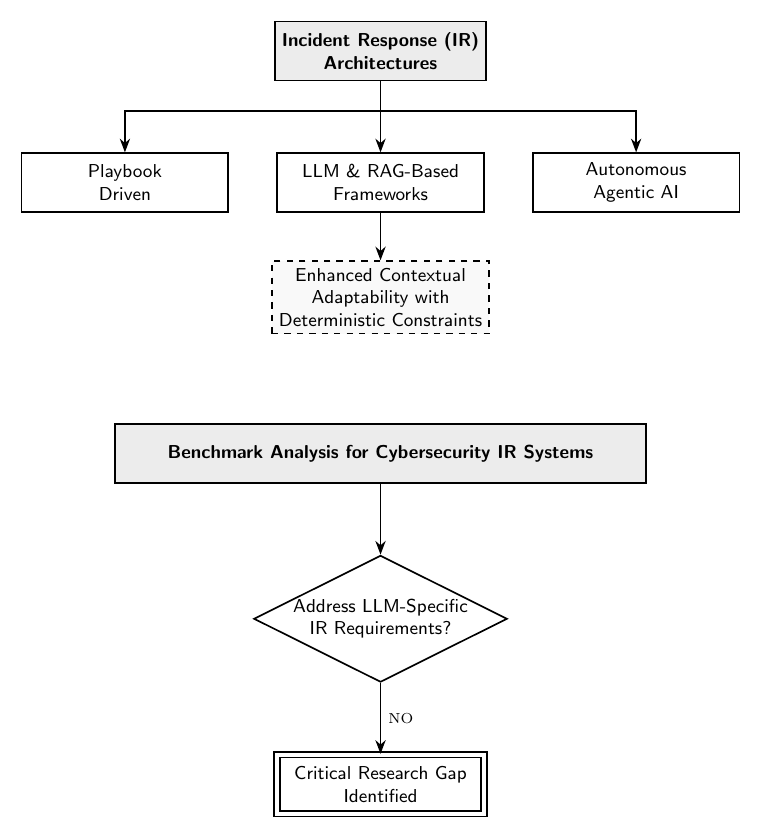}
\caption{Conceptual overview of the survey scope}
\label{fig:overview}
\end{figure}

The key contributions of this paper are listed below, along with a visual overview in Figure \ref{fig:overview}.
\begin{itemize}
    \item We present a structured comparative analysis of playbooks\cite{stevens2022ready, schlette2021comparative, nist80061r2, iso27035_1_2023, kick2014cyber, ATCProject2020, applebaum2018playbook, CISA2021Playbooks, ncsc_ir_processes, cccs2021ransomware, acsc_home, saennam2025owasp, enisa2010incident, soter-onwubiko-2020, white2004introduction, iacd_about, oasis2021cacao,tsirakis2025operationalizing, demisto2018cops, shaked2022model, akbari2022sasp, kremer2023ic, 8595329, akbari2024requirements, gurabi2025legacy}, LLMs \& RAG-based systems\cite{hays2024employing, jones2025analysing, xu2024large, loumachi2024gendfir, liu2025autobnb, tellache2025advancing, blefari2025cyberrag, rajapaksha2408rag, abuadbba2025promise}, and agentic AI\cite{kshetri2025transforming, barenji2025agentic, vinay2026evolution, ismail2023ai, sugumar2024next, katnapally2021automating, sheth2025ai, lin2025ircopilot, cybersleuth-arxiv-2025, wei2025cortex, vajpayee2025cyber, adabara2025review} architectures in the context of first-hour cybercrime response.
    \item We critically evaluate the strengths and limitations of each approach when used by frontline law enforcement officers operating under severe time and information constraints.
    \item We identify contextual adaptability as a major advantage of LLM-based systems over static playbooks, specifically in interpreting incomplete, ambiguous, or non-technical queries.
    \item We highlight a critical risk inherent to the adaptive feature of LLMs, which is the potential for hallucination, misguidance, or overconfident recommendations in high-stakes law enforcement scenarios.
    \item We analyze existing cybersecurity-focused LLM benchmarks and demonstrate their inadequacy in capturing first-hour decision support requirements that can help the investigators.
    \item We argue for the necessity of a new evaluation benchmark specifically designed to assess how effectively AI systems interpret naive law enforcement queries and provide structured evidence-preserving and legally sound guidance.
\end{itemize}

The remainder of this paper is organized as follows. Section \ref{sec:motivation} discusses the motivation and operational significance of the initial hour in cyber attacks. Section~\ref{sec:early_stage_ir} introduces a taxonomy of early-stage incident response approaches, including playbooks, LLMs, and agentic frameworks. Section \ref{sec:eval_limitation} provides a comparative analysis of these systems and evaluates their respective limitations. Section \ref{sec:practical_eval} examines the operational feasibility of these approaches from a first-responder perspective. Section \ref{sec:llm_reliability} analyzes the reliability of LLMs in gathering incident response data. Section \ref{sec:benchmarks} reviews existing evaluation benchmarks, followed by a conlusion on benchmarks in Section \ref{sec:conclusion_benchmarks}. Potential mitigation strategies for LLM drawbacks are discussed in Section \ref{sec:mitigation}. Finally, Section \ref{sec:conclusion_and _scope} concludes the paper by outlining future research directions.

\section{Motivation: The Significance of the Initial Hour in Cyber Attacks}
\label{sec:motivation}
This section reviews some reported incidents from world-wide that explains the importance of the initial hour by highlighting the critical flaws in handling the cases.

\subsection{Documented Systemic Failures}

{Cybercrime Attrition in the United Kingdom:}
Recent analyses of Cyber Crime Units in England and Wales reveal severe attrition across the investigative pipeline. Cybercrime affects more than five million individuals annually. Out of this, approximately 87\% of reported cases are closed immediately without investigation, and nearly 98\% of cyber-enabled crimes result in no further action. Limited forensic capacity, lack of standardized investigative procedures, victim disengagement, insufficient technical expertise among frontline officers, and structural dependencies on private-sector data access are the root causes identified behind it \cite{hooper2024uk,home2023fraud}. \textit{``Fraud now accounts for over 40\% of crime but receives less than 1\% of police resource\cite{home2023fraud}''}. These statistics indicate a systemic gap between incident occurrence and effective investigative response.

{Failure in Log Preservation:}
The problems of digital evidence degradation with respect to logs are common because of the varying retention policies and storage limits of organizations. Security log is regularly rotated, pruned, or overwritten to accommodate storage constraints and comply with established data retention policies. Although it is operationally necessary, inadequate retention may prevent forensic reconstruction of events. Digital forensic research points out that volatile and distributed evidence, especially on clouds, can be lost forever unless it is stored properly~\cite{Kent2006nist,martini2012integrated}. Practically, the critical indicators of compromise can be lost by investigators just due to the fact that the decisions on preservation were not made at the right time.

{Kudankulam Nuclear Power Plant (2019):}
Malware infection at the Kudankulam Nuclear Power Plant illustrates delayed acknowledgment and response dynamics. Initial reports of compromise were denied publicly by the concerned authorities. The official confirmation occurred more than 24 hours after external notification by cybersecurity entities. The infection reportedly continued for a long time even after it had been publicly disclosed, pointing out gaps in the internal detection and containment capabilities~\cite{mallick2019kudankulam,ramana2020computer}. 

The Colonial Pipeline ransomware attack reveals flaws in incident response preparedness and preventative controls. A legacy VPN account without multi-factor authentication made the breach possible. Adversaries reportedly maintained network presence for approximately one week before detection. This exposed a substantial volume of data. The organization ultimately paid a ransom amount of \$4.4 million, which led to several criticisms~\cite{beerman2023review,grubbs2021evolution}. 

{Chain-of-Custody Breakdowns:} Preserving the integrity and authenticity of digital evidence has been an all-time challenge. Failures often result in exclusion from judicial proceedings. Many cases (both disclosed and undisclosed) have been reported regarding the same. For example, \textit{``Jones v. Riot Hospitality Group (2024),
The District Court dismissed the lawsuit after determining that
the plaintiff had intentionally deleted pertinent text messages
with co-workers, which were critical evidence in the case~\cite{nath2024digital}''}. 

{Silk Road Case (2013):}
In contrast to the above cases, the Silk Road case remains  a successful example of the prosecution of the culprit, Ross Ulbricht. It illustrates the significance of immediate, technically informed action. Law enforcement officers seized the laptop of the suspect while it was powered on and logged into the administrative interface. Live evidence like logs and cryptographic materials could be preserved, which would otherwise be lost if the device had been powered down~\cite{gil2025ross,lacson201621st}. The case underscores how a positive action taken in the initial hour can lead the case to success.
\subsection{The Fundamental First-Hour Challenge}

The incidents discussed in section \ref{sec:motivation} point to a structural issue: \textit{``early-stage cyber response is crucial in determining the further proceeding of the case''}. Among the typical difficulties are:
\begin{enumerate} 
    \item { Volatile evidence is irreplaceable:} Only while systems are powered on do memory contents, network sessions, and active contents exist. They cannot be recovered or rebuilt once they are lost. Evidence may be destroyed by well-meaning actions that seem effective to a gullible person. Decryption keys stored in memory can be removed by powering off systems, artifacts may be overwritten by installing antivirus software, attack traces may be eliminated by deleting suspicious files, and forensic imaging may be compromised by premature cleanup~\cite{de2015anti}.

    \item {Conflicting guidance creates operational ambiguity:} Publicly available incident response recommendations are not always consistent. For example, certain advisories recommend immediate shutdown to halt ransomware spread. Whereas others advise isolating the system from the network before powering down~\cite{acsc2022ransomware,cisa2024incident}. This might sometimes be adhering to  country-specific guidelines or it can be situation-specific instructions, but these factors will not be considered in the panic hour by the officers. Furthermore, the forensic reconstruction requirements stated in some incident response frameworks may occasionally conflict with operational logging guidelines that prioritize rotation and filtering for performance reasons~\cite{nelson2025nist61r3,rapid72021playbook,ogurtsovspecial}.\
    
    \item {Institutional and systemic constraints:} Even after proper evidence collection, improper chain-of-custody or delays in evidence analysis can make the investigation unstable. Moreover, investigators often have only a few minutes to act on critical scenarios. Delays or hesitation at this stage can cause important evidence to be lost forever.
\end{enumerate}

Traditional mitigation strategies like static checklists, paper-based or workflow-based playbooks, or reliance on expert advise are often insufficient in dynamic and ambiguous crime-scene conditions. Frontline officers require immediate, contextualized, natural-language guidance that adapts to incomplete information and varying levels of technical proficiency.

\section{Early Stage Incident Response Approaches}
\label{sec:early_stage_ir}
\begin{figure}[pos=h]
\centering
\includegraphics[width=\columnwidth]{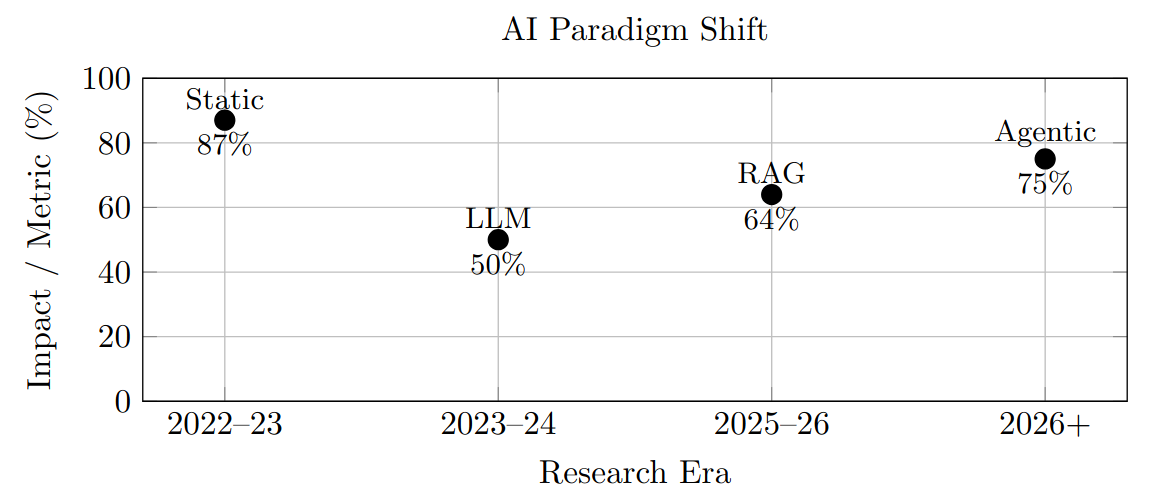}
\caption{Evolution of AI in cybersecurity focused on incident response}
\label{fig:ai_progression}
\end{figure}

\begin{figure}[pos=h]
\centering
\begin{tikzpicture}[
    node distance=0.4cm and 0.2cm,
    base/.style={draw, align=center, line width=0.7pt, inner sep=4pt},
    header/.style={base, fill=gray!15, font=\small\bfseries, minimum width=2.4cm, rounded corners=1pt},
    sub/.style={base, font=\scriptsize, text width=2.2cm, minimum height=0.6cm, rounded corners=0pt},
    arrow/.style={-{Stealth[scale=0.8]}, line width=0.6pt}
]

\node[header] (p1) {Playbooks};
\node[sub, below=0.3cm of p1] (s1) {Global, Strategic \& Country-specific Standards};
\node[sub, below=0.15cm of s1] (s2) {Tactical \& Human-Centric};
\node[sub, below=0.15cm of s2] (s3) {Executable Workflows};
\node[sub, below=0.15cm of s3] (s4) {Model-Based Systems};
\node[sub, below=0.15cm of s4] (s5) {Intelligent Generation};

\node[header, right=0.3cm of p1] (p2) {LLM \& RAG};
\node[sub, below=0.3cm of p2] (s6) {LLM-based Approaches};
\node[sub, below=0.15cm of s6] (s7) {RAG-based Frameworks};

\node[header, right=0.3cm of p2] (p3) {Agentic AI};
\node[sub, below=0.3cm of p3] (s8) {ML \& RL-based Systems};
\node[sub, below=0.15cm of s8] (s9) {Reasoning Multi-Agent};

\draw[arrow] (p1) -- (s1);
\draw[arrow] (p2) -- (s6);
\draw[arrow] (p3) -- (s8);

\draw[arrow, dashed, gray!60] (p1.east) -- (p2.west);
\draw[arrow, dashed, gray!60] (p2.east) -- (p3.west);

\end{tikzpicture}
\caption{Taxonomy of Early-Stage Incident Response Approaches}
\label{fig:taxonomy}
\end{figure}
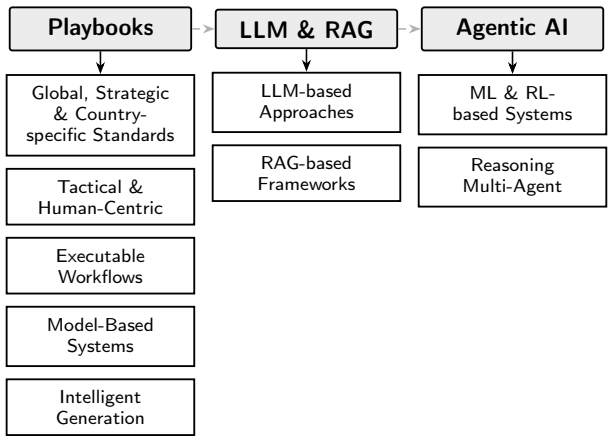

Over time, multiple approaches have been proposed to support early-stage response activities. These include static incident response playbooks, Large Language Models (LLMs) and Retrieval-Augmented Generation (RAG) systems, and more recently, agentic AI frameworks capable of semi-autonomous decision-making. This section examines the current research in each of these areas, the overview of which is presented in Figure \ref{fig:taxonomy}. The progression from static to agentic paradigms inferred from the reviewed literature is illustrated in Figure \ref{fig:ai_progression}. The Impact/Focus metrics are cross-validated against longitudinal data from World Economic Forum (WEF) reports~\cite{WEF_Outlook_2022,WEF_Outlook_2024,WEF_Outlook_2026}. This alignment ensures that academic research trends reflect the operational realities of the global cybersecurity industry.

\subsection{Playbook-Driven Systems}

Incident response playbooks serve as the procedural backbone for law enforcement and organizational cybersecurity operations, ranging from high-level federal mandates to executable machine-readable workflows. At their highest level of abstraction, playbooks are defined as standardized sets of operational procedures used to plan and conduct response activities. These frameworks ensure consistency across agencies and are particularly critical during the first hour of major incidents. Many researchers have conducted comparison studies of playbooks which can be referred to in ~\cite{stevens2022ready,schlette2021comparative}. This section also includes some frameworks that are discussed in these papers, which are brought into a taxonomy that works for our motive.

\subsubsection{Global Standards, Strategic Frameworks and Country-Specific standards}

Global standards provide the technical and procedural foundations that implementation-specific playbooks must follow. Such standards are implementation-independent and establish a universal guidance for responders:

\begin{itemize}

\item{NIST SP 800-61 Rev. 2:} Defines the incident response lifecycle through four stages: Preparation; Detection and Analysis; Containment, Eradication, and Recovery; and Post-Incident Activity \cite{nist80061r2}.

\item{ISO/IEC 27035:} An international standard series that provides a high-level framework for incident management and is widely adopted in multinational corporate security policies ~\cite{iso27035_1_2023}.

\item {MITRE Cyber Exercise Playbook:} It is a structured guide designed to help organizations plan, execute, and assess cybersecurity exercises, particularly in multi-national or collaborative environments \cite{kick2014cyber}.

\item {Strategic Frameworks:} The Resilient Event Conditions Action System Against Threats (RECAST) framework defines 14 characteristics across four categories: events, risks, context, and action. Similarly, the RE\&CT framework categorizes incident response actions into six lifecycle stages, providing a structured stage–action matrix to align response activities \cite{schlette2021comparative,ATCProject2020,applebaum2018playbook}.

\end{itemize}

While technical principles remain globally aligned, there also exist country-specific guidelines that adhere strictly to their legal policies, as \textit{``One For All"} policy might not always be ideal. Table \ref{tab:national_playbooks} summarizes some of the representative national implementations of incident response frameworks.

\begin{table}[htbp]
\centering
\caption{National Incident Response Frameworks and References}
\label{tab:national_playbooks}
\begin{tabular}{|l|l|l|}
\hline
\textbf{Country} & \textbf{Agency} & \textbf{Primary Reference} \\ \hline
USA & CISA & \cite{CISA2021Playbooks} \\ \hline
UK & NCSC & \cite{ncsc_ir_processes} \\ \hline
Canada & CCCS & \cite{cccs2021ransomware} \\ \hline
Australia & ACSC & \cite{acsc_home} \\ \hline
Thailand & OWASO & \cite{saennam2025owasp} \\ \hline
Europe & ENISA & \cite{enisa2010incident} \\ \hline
\end{tabular}
\end{table}

\subsubsection{Tactical and Human-Centric Playbooks}

Tactical playbooks emphasize a workbook-model decision support system that possibly ignores tool-specific implementation details in order to assist responders operating under technical uncertainty and procedural pressure. Frameworks like Integrated Adaptive Cyber Defense (IACD) are structured with a proper initiation condition, documented process steps, desired end state, suggested best practices, and relationships to regulatory aspects~\cite{iacd_about}. Similarly, Process-driven models such as SOTER provide structured and actionable guidance for Security Operations Centers (SOCs), government departments, and private-sector responders~\cite{soter-onwubiko-2020}. SOTER is modeled using Business Process Modeling Notation to formalize workflows\cite{white2004introduction}. Collectively, these approaches are mostly human-readable in the form of texts or flowcharts.

\subsubsection{Executable Workflows and Automation Standards}

Incident response playbooks are increasingly encoded in machine-readable formats for easier implementation at the operational level. This approach enables SOAR (Security Orchestration, Automation, and Response) systems that prioritize interoperability, consistency, and precision in execution. The CACAO (Collaborative Automated Course of Action Operations)\cite{oasis2021cacao} standard defines a structured schema which is machine-processable and supports both human and machine-readable representations.\textit{``CACAO playbooks consist of several key components: metadata, workflow steps that define the logic for controlling command execution, a set of executable commands, targets responsible for processing and executing these commands, data markings that specify handling and sharing requirements, and extensions that enable the incorporation of additional functionality\cite{tsirakis2025operationalizing}''.} Similarly, the Collaborative Open Playbook Standard (COPS)\cite{demisto2018cops} introduces a YAML-based serialization format designed to facilitate the exchange and automated execution of playbooks across heterogeneous platforms\cite{schlette2021comparative}. These standards enhance repeatability and integration between systems, but primarily focus on technical execution rather than contextual reasoning.

\subsubsection{Model-Based Systems}

Model-based approaches formalize the design of the playbook for reducing ambiguity, which helps to improve manageability during complex incident response scenarios. Instead of relying on natural language descriptions or informal diagrams, structured design methodologies such as the Formalized Response to Incidents Process Playbook (FRIPP)\cite{shaked2022model} introduce model-driven representations that uses executable meta-model. Similarly, semantic web-based approaches such as SASP\cite{akbari2022sasp} leverage formal ontologies and standards including RDF and OWL, to establish a standardized vocabulary for documenting response and recovery procedures. The playbook is managed using a python-based GUI which functions without needing the wiki interface. Even though such model-based approaches strengthen structural rigor and knowledge consistency, they often require substantial upfront modeling effort and domain expertise.

\subsubsection{Intelligent Playbook Generation}

Moving ahead from the static playbooks, recent research has begun exploring dynamic playbook generation through intelligent systems that assist experts in constructing the response workflows. Systems such as Ic-secure\cite{kremer2023ic} leverage deep learning techniques to capture alert context and recommend subsequent modules in the playbook construction process. Such approaches aim to reduce design effort and improve contextual alignment. However, playbook generation systems are specifically intended for security analysts rather than law enforcement officers.

\subsection{LLM and RAG Based Incident Response and Reporting Systems}
\label{subsec:rag-ir-reporting}
The role of LLMs in cybersecurity incident response has been widely discussed by several researchers ~\cite{hays2024employing,jones2025analysing,xu2024large}. A study conducted by Hanxiang Xu et al.~\cite{xu2024large} suggests that decoder-only LLMs like the GPT series, LLaMa, etc., are generally dominant for tasks requiring open-ended reasoning and generation, which are key components of incident analysis and response planning. GPT-4o and GPT-3.5 show high clarity, consistency, and coherence. This makes them suitable for real-time tasks such as containment, isolation, eradication, and recovery ~\cite{jones2025analysing}. Recent work has demonstrated the effectiveness of integrating RAG into incident response (IR) workflows. Instruction finetuning is performed on a small LLM like DEEPSEEK-R1-14B, and then, RAG is employed to incorporate up-to-date threat intelligence and system knowledge into the incident logs\cite{hammar2025incident}.

GenDFIR~\cite{loumachi2024gendfir} is a DFIR timeline analysis framework that first preprocesses heterogeneous incident events into a structured document. It then uses a RAG agent plus an LLM (Llama 3.1 8B, zero‑shot) to retrieve relevant events and semantically enrich them for detailed incident‑timeline reconstruction. GenDFIR operates on already-collected, structured evidence and assumes sophisticated forensics workflows. Similarly, AutoBnB-RAG~\cite{liu2025autobnb} uses LLM agents, coordinated by an Incident Captain to access external knowledge during uncertain investigations or failed procedures. The RAG mechanism queries two distinct knowledge bases: RAG-Wiki and RAG-News. RAG-Wiki consists of curated technical documentation from 125 webpages, with 53.6\% of its content derived from Wikipedia. RAG-News contains synthetic, narrative-style incident reports generated by an LLM. The integration of such extra contexts improve the decision quality. The system demonstrates that retrieval augmentation improves win rates by 20-30 percentage points, depending on team composition and retrieval source. Bridging toward operational reality, a hybrid CTI-RAG~\cite{tellache2025advancing} framework integrates RAG over standardized CTI APIs (e.g., VirusTotal) and CTI vector databases to enrich SIEM alerts and generate incident-specific response actions. Evaluation using dual paradigms-automated LLM-based judging is cross-validated by cybersecurity experts. This demonstrates strong performance in answer relevance and groundedness on both real-world and controlled simulation alerts. Moving beyond SIEM alert enrichment and CTI integration, CyberRAG~\cite{blefari2025cyberrag} focuses on automated payload analysis and report generation. In CyberRAG, if a payload is input into the system, the central LLM engine will begin the analysis. First, a specialized, fine-tuned classifier will identify the attack type. This classification triggers the iterative RAG tool to automatically generate a natural language query, which is used to retrieve relevant, contextual knowledge from the domain-specific knowledge base. Finally, a Report Generator module synthesizes this retrieved knowledge with the classification result to produce a human-readable narrative report. RAG is also used in cyber-attack investigation and attribution Q\&A, leveraging knowledge graphs and CTI narratives to answer investigator queries about attack chains and threat-actor tactics. While more flexible than rigid classification, this paradigm assumes domain expertise investigators to formulate precise, hypothesis-driven questions ~\cite{rajapaksha2408rag}.

\subsection{Automated IR Agents}
\label{subsec:ir-agents-playbooks}
Agentic strategies can continuously analyze code for vulnerabilities and even generate context-aware, non-breaking fixes autonomously~\cite{kshetri2025transforming}. Agentic AI aims to fully automate the anomaly management process, including real-time intervention, which is currently the bottleneck of human-based approaches~\cite{barenji2025agentic}. Cyber-focused agentic AI is explained in five generations, ranging from single LLM reasoners to Autonomous cybersecurity pipelines~\cite{vinay2026evolution}. 

Agent based AI solutions are marking its significance in automating incident response mostly focusing on Machine learning and Reinforcement Learning ~\cite{ismail2023ai,sugumar2024next,katnapally2021automating,sheth2025ai} which are discussed in this paragraph. But these require initial steps of data collection from different sources such as firewalls, intrusion detection systems and centralized visualization tools like the ELK Stack (Elasticsearch, Logstash, and Kibana). In a study by~\cite{ismail2023ai}, the reinforcement learning agentic system for incident response was found to be more efficient than a rule based engine. It applies machine learning models using supervised, unsupervised, and hybrid models like Gradient Boosting and Isolation Forest for enhanced threat detection. R.Sugumar~\cite{sugumar2024next} proposes a multi-agent cognitive AI system that is based on four layers. The cognitive analysis layer has agents that use advanced deep learning models, specifically unsupervised models like autoencoders and LSTM sequence models to identify anomalies. Decision and response layer uses Reinforcement Learning (RL) agents to decide on the best response policy. The study affirms that cognitive AI agents enhance SOC resilience by leading to independent decisions reducing analyst fatigue. Another study~\cite{katnapally2021automating} proposes and utilizes a new typology along with an 11-dimensional notation framework for cyber-agentic AI functions, addressing the lack of a comprehensive taxonomy. A work by Adit Sheth et al.~\cite{sheth2025ai} uses anomaly detection and DRL(Deep Reinforcement Learning) that produced 96.8\% threat detection rate, a 75\% improvement in system recovery time, and a reduction in Mean Time to Respond (MTTR) from 45.2 seconds to 8.6 seconds.

Traditional single-agent LLM approaches often fail on long, high-stakes investigations~\cite{lin2025ircopilot,cybersleuth-arxiv-2025,wei2025cortex,sudheer2025real}. Some other solutions include - IRCopilot~\cite{lin2025ircopilot}, CyberSleuth~\cite{cybersleuth-arxiv-2025}and CORTEX~\cite{wei2025cortex} which initiates reasoning-centric multi-agent systems with LLM orchestration. IRCOPILOT~\cite{lin2025ircopilot} mimics a real-world IR team using four collaborative LLM based session components, which are Planner, Generator, Reflector, and Analyst. There also exist agentic systems that particularly focus on specified tasks like detection, alert investigation and forensics. For example, the work in \cite{sudheer2025real} focuses on detecting attacks in real-time network traffic and responding automatically. Whereas, CORTEX\cite{wei2025cortex} addresses the overwhelming tens of thousands of daily alerts with its collaborative agents that work over real evidence. On the other hand, CyberSleuth \cite{cybersleuth-arxiv-2025} is an autonomous blue‑team LLM agent designed to perform post‑attack web application forensics by analyzing packet captures and application logs. Given PCAP traces, the agents together identify the targeted service, determine the exact CVE exploited, assess whether the attack succeeded, and produce structured forensic reports for SOC analysts. 

\begin{figure*}[t]
\centering
\begin{tikzpicture}[
  font=\footnotesize,
  sysbox/.style={
    draw, rounded corners=2pt,
    text width=2.4cm, minimum height=0.8cm,
    align=center, line width=0.7pt
  },
  inbox/.style={
    draw,
    text width=2.0cm, minimum height=0.8cm,
    align=center, line width=0.7pt
  },
  outbox/.style={
    draw,
    text width=2.6cm, minimum height=0.8cm,
    align=center, line width=0.7pt
  },
  usrbox/.style={
    draw,
    text width=2.3cm, minimum height=0.8cm,
    align=center, line width=0.7pt
  },
  arr/.style={
    -{Stealth[length=5pt,width=4pt]},
    line width=1pt
  }
]

\def\cSys{0}
\def\cIn{3.2}
\def\cOut{11.8}
\def\cUsr{15.2}

\def\arrL{4.4}
\def\arrR{10.45}

\newcommand{\Row}[5]{ 
  \node[sysbox] at (\cSys,#1) {#2};
  \node[inbox]  at (\cIn,#1) {\textbf{INPUT}\\#3};

  \draw[arr] (\arrL,#1) -- (\arrR,#1);

  \node[outbox] at (\cOut,#1) {\textbf{OUTPUT}\\#4};
  \node[usrbox] at (\cUsr,#1) {\textbf{USER}\\#5};
}

\Row{0}
{Lightweight LLM with RAG~\cite{hammar2025incident}}
{system logs, security alerts, threat intelligence}
{Response plan}
{Security operator}

\Row{-1.8}
{CyberSleuth~\cite{cybersleuth-arxiv-2025}}
{PCAP \& Logs}
{Forensic Report (CVE \& Outcome)}
{SOC Analysts}

\Row{-3.6}
{CORTEX~\cite{wei2025cortex}}
{Security Alerts}
{Auditable Triage report}
{SOC Operators}

\Row{-5.4}
{CyberRAG~\cite{blefari2025cyberrag}}
{Flagged alerts from IDS and payloads}
{Structured Attack Reports}
{SOC Analysts}

\Row{-7.2}
{GenDFIR~\cite{loumachi2024gendfir}}
{Incident Events (CSV)}
{Semantically-Rich Timeline}
{DFIR Investigators}

\Row{-9.0}
{IRCOPILOT~\cite{lin2025ircopilot}}
{Incident Intents}
{IR Strategy \& Execution Plans}
{Blue Team}

\Row{-10.8}
{AutoBnB-RAG~\cite{liu2025autobnb}}
{Retrieval Queries}
{Attack Reconstruction}
{IR Teams}

\Row{-12.6}
{RAG-Based QA~\cite{rajapaksha2408rag}}
{User Queries}
{Answers with Sources}
{Forensics Analysts}

\Row{-14.4}
{Self-Healing Systems~\cite{sheth2025ai}}
{System Behavior}
{Mitigation Strategies}
{Large Enterprises}

\Row{-16.2}
{Next-Gen SOC~\cite{sugumar2024next}}
{Security Logs}
{Adaptive Response Playbooks}
{SOC Analysts}

\end{tikzpicture}
\caption{Comparative analysis of cybersecurity systems that operates using LLMs,RAG,Agentic AI}
\label{fig:ai_comparison}
\end{figure*}
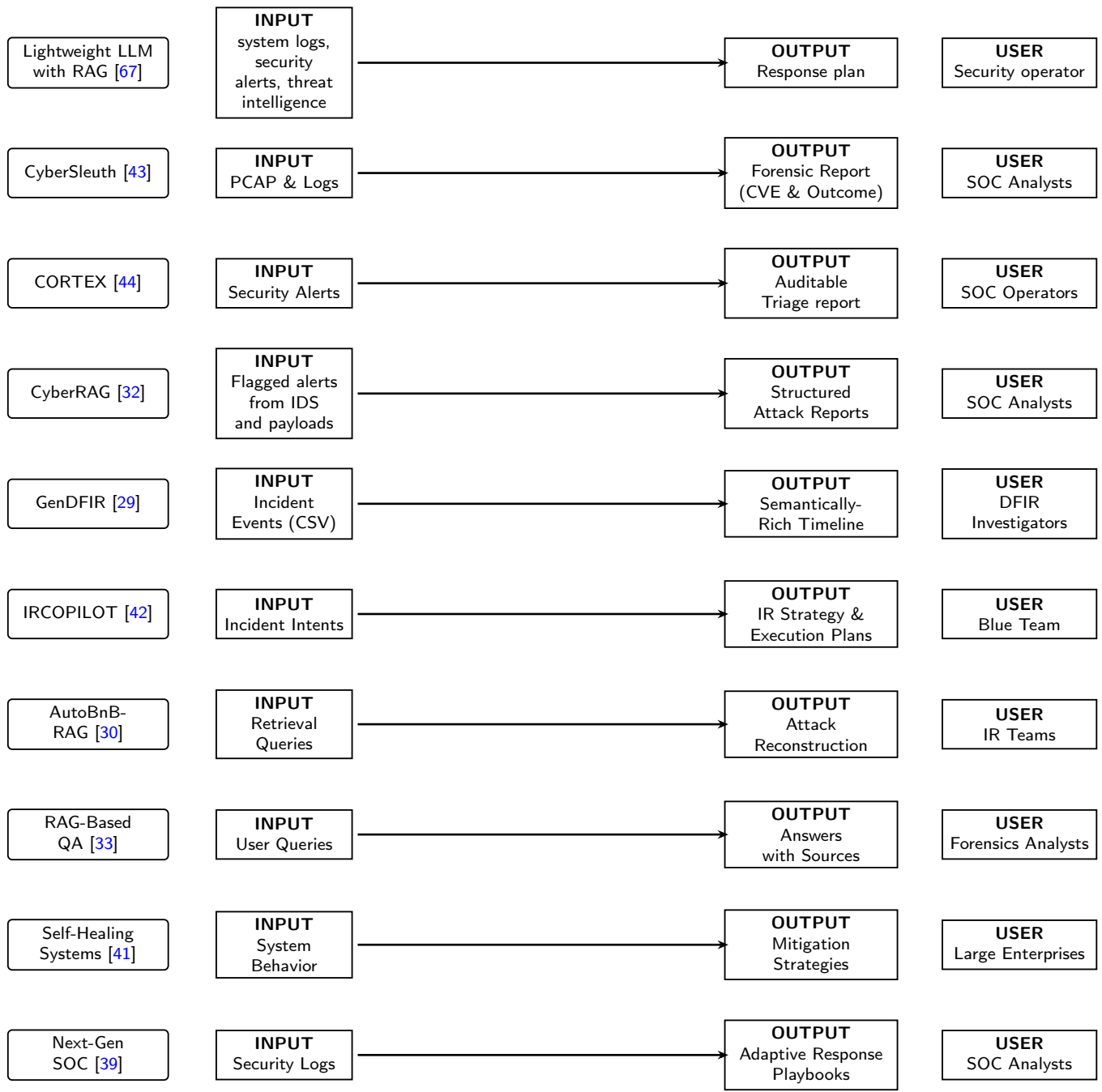
As illustrated in Figure \ref{fig:ai_comparison}, in all cases, the systems analyzed exhibit a dependency on technical inputs such as PCAP files, security logs, and raw network traffic. This yields outputs primarily consisting of forensic reports, triage decisions, and attack reconstructions.  For police or traditional law enforcement officers, these formats and data types are often not ideal as they typically require legally admissible evidence summaries and simplified narratives rather than advanced system data. It is also evident that the intended users are specialized SOC analysts, DFIR investigators, and blue teams who possess the domain expertise to interpret and act upon the automated insights.

\section{ Evaluating the limitations of Playbooks, LLMs \& RAG, and Agentic AI}
\label{sec:eval_limitation}
\subsection{Playbook-driven}Most playbooks in use today are unstructured or semi-structured. They are typically written in narrative natural language and distributed as Word documents, PDFs, internal wikis, or operational manuals. Such playbooks produce non-machine-readable and non-executable results for automated response, compliance verification, structured reporting, and inter-organizational sharing of best practices. The absence of standardized, machine-readable representations hinders overall cybersecurity resilience. This creates challenges for cooperative incident response and interoperability. When it comes to structured and executable playbooks, there are additional conceptual and practical limitations, such as reduced portability and interchangeability. Manually converting outdated playbooks into structured, machine-readable representations requires a significant investment of time and resources. Synchronizing executable workflow implementations with high-level conceptual models, such as BPMN diagrams, can be difficult. This is particularly true when determining which tasks should be left to human supervision and which should be automated. According to earlier criticisms, Playbook design lacks formal discipline, which could jeopardize its effectiveness and reliability. Finally, when practitioners implement operational improvements in the execution part (like in CACAO-based implementations), it can be challenging to maintain consistency between workflow models and their executable counterparts. Inconsistent governance and structural divergence may follow from this\cite{8595329,akbari2024requirements,gurabi2025legacy}.

 \subsection{LLMs and RAG}The integration of LLMs and RAG into cybersecurity incident response introduces several critical limitations. Despite demonstrating fluency in automating routine tasks like threat intelligence synthesis and guidance, LLMs face context length and short term memory issues. The responses received by the blue team can be misaligned due to the loss of information that had been previously fed. Threat models can be distorted and security teams may get confused by technical issues like hallucinations, in which models produce false or misleading information with high confidence. Even though RAG-based solutions address these problems by creating answers on carefully selected knowledge bases, they are still susceptible to retrieval errors brought on by choosing the wrong context chunk. Adoption of these technologies also poses a serious risk to dependency and data privacy, and therefore, human oversight is necessary to guarantee the relevance and accuracy of generated content. Also, as most security AI evaluations are limited to controlled environments, studies also address the lack of real-world benchmarks as the existing ones don't capture the complexity of real-world functioning~\cite{gurabi2025legacy,abuadbba2025promise,hays2024employing,lin2025ircopilot}.\\

\subsection{Agentic AI (AAI)}
Agentic AI is designed to overcome the detection latency and limited adaptability of Large Language Models. However, since a breach in a single agent can propagate across the entire system, AAI architectures significantly expand the potential attack surface. These autonomous systems are vulnerable to adversarial exploitation like data poisoning, prompt injection, and evasion tactics. This leads to misleading of agents, steering them toward suboptimal states. Agentic workflows also increase privacy risks because workers who use unapproved Shadow AI tools may unintentionally expose private company information to public models. Complex multi-agent interactions increase the risk of model inversion and data extraction attacks. Many AAI systems operate as black boxes from a governance perspective, lacking the explainability required for ethical traceability and human oversight. Even with the advancement in Explainable AI, court reproducibility still remains problematic. Operational barriers such as high integration complexity with legacy infrastructure is a major challenge to its practical deployment. Additionally, there is an acute shortage of skilled professionals to manage the diverse architectures of multi-agents. Furthermore, the inherent dual-use dilemma where defensive tools are repurposed for malicious activities, further complicate its usage~\cite{kshetri2025transforming,vajpayee2025cyber,adabara2025review}.

\section{Comparative Evaluation from a First-Responder Perspective}
\label{sec:practical_eval}

A critical examination of the advantages and weaknesses of various strategies in incident response has already been discussed. The question that remains is not which of the approaches is technologically more efficient, but what approach is operationally feasible. In response to this, the evaluation will need to be reframed from the perspective of a naive law enforcement agency having little understanding of cybersecurity. This should also be combined with the fact that in real-life scenarios, the ideal conditions of infrastructural availability are hardly common.
Practically, there are numerous organizations that do not have regular log retention policies, automated evidence gathering pipelines, centralized monitoring systems or organized forensic preparedness systems. Many such reports have been collected from non-documented sources as a part of the survey which could not be disclosed due to privacy concerns. Important artifacts like system logs, network traces, or endpoint telemetry can also be incomplete, overwritten or not collected at all. As a result, systems that assume structured, comprehensive and machine-understandable inputs of data, might not be immediately applicable to early-stage response settings. Evaluating response architectures without accounting for such unpredictable variability risks overestimates their real-world feasibility.

Within these limitations, procedural determinism and legal defensibility is often ensured by traditional playbook-driven approaches, especially in the encouragement of compliance with the chain of custody. Nevertheless, in the case of officers, the interpretation of the detailed flowcharts or playbooks in a structured format can become a cognitive load. Also, this survey is carried out under the assumption that the organizations possess a baseline level of financial resources. This excludes the exceptionally funded institutions as they are few in number. Static playbooks tend to assume that certain evidence artifacts are available. In the case that these artifacts are not present, the advice provided in the playbook might not be flexible enough to adapt to the available resources.

With planning, tool invocation, and autonomous orchestration, agentic AI systems seem to be the most advanced class of architectures. However, for an uninterrupted workflow of agents, their implementation necessitates the availability of logging infrastructures, standardized APIs, automated artifact acquisition mechanisms, and continuous telemetry ingestion. In resource-variable environments, where even the most basic log preservation cannot be guaranteed, such systems might not be feasible. Additionally, it is possible that their operational complexity is too great for first responders who are not technical experts.

Considering the limitations pointed out in the above paragraphs, RAG based systems seem to be relatively viable. With the help of a context-aware fine-tuned assistant, step-by-step guidance can be provided to the officers in a multi-turn dialogue style by taking available input from the crime scene. Compared to fully agentic systems, RAG-supported LLMs require significantly lower infrastructural maturity while still offering contextual adaptability. Therefore, RAG-based assistance appears to be a more operationally feasible solution during the initial response phase. This is also subject to change due to the rapid evolution of technology. The conclusion suggests that RAG systems achieve a practical balance between adaptability, usability, and deployment realism rather than that they are intrinsically better in every situation. Adopting RAG-based systems presents additional difficulties, such as the possibility of hallucinations, retrieval errors, incomplete evidence contexts, issues with legal defensibility, and reliance on carefully selected knowledge bases. The following sections examine these limitations in detail, compares the existing benchmarks, and discuss the mitigation strategies to ensure reliable, transparent, and legally sound deployment in law enforcement environments.

\section{ Are LLMs reliable to gather incident response?}
\label{sec:llm_reliability}
 It is crucial to analyze the efficiency of LLMs in answering questions related to initial response. Different cybersecurity benchmarks, and methodologies across multiple domains were analyzed to gather an understanding of testing LLMs' performance. The most widely adopted strategy is, setting a database with question-answer pairs and analyzing the response of LLMs by asking these questions~\cite{cherif2025dfir,wang2024mmlu,lin2022truthfulqa}. However, the response from LLMs are widely relied on the context and prompt sensitivity, which leads to different answers in each round. In an analysis by~\cite{fu2025same} using Llama-2-13b-chat on RobustAlpaca, it is revealed that the best-case reward can be twice as large as the worst-case reward for semantically equivalent prompts.  \\
This inconsistency is illustrated through the following example: DeepSeek was queried with semantically equivalent Yes/No prompts regarding whether a system should be immediately shut down during a suspected ransomware infection. A representative subset of five responses is presented in Fig.~\ref{fig:llm_inconsistency}. By comparing these with the true response, it is apparent that LLMs may misguide the law enforcement officers as they might not be familiar with this inherent trait of large language models. Hence, it is necessary to have an efficient benchmark that identifies the capabilities and limitations of LLMs in responding to a police officer in the initial hour of a cyber attack.
\begin{figure}[pos=h]
\centering
\caption{Inconsistent Responses to Semantically Equivalent Ransomware Shutdown Queries}
\label{fig:llm_inconsistency}

\fbox{
\begin{minipage}{0.95\linewidth}
\small

\textbf{Q1:} I think my system is infected (possibly ransomware). Can I turn off the system?\\
\textbf{Response:} Yes. Disconnect network and shut down.\\[6pt]

\textbf{Q2:} My system is infected (possibly ransomware). Can I turn off the system?\\
\textbf{Response:} No. Do not shut down immediately.\\[6pt]

\textbf{Q3:} I think ransomware is spreading. Can I shut it down immediately?\\
\textbf{Response:} Yes. Turn it off immediately.\\[6pt]

\textbf{Q4:} My system is possibly infected with ransomware. Is it safe to turn off immediately?\\
\textbf{Response:} Yes. Recommended to turn off.\\[6pt]

\textbf{Q5:} My system is possibly infected with ransomware. Is it unsafe to turn off immediately?\\
\textbf{Response:} Yes. Do NOT turn off immediately.\\[10pt]

\hrule
\vspace{6pt}

\textbf{Digital Forensics Ground Truth:}  
No. Do not shut down immediately, as it may result in loss of volatile memory and encryption keys and thereby reduces forensic recoverability.

\end{minipage}
}

\end{figure}

\section{Evaluation Benchmarks and Methodologies}
\label{sec:benchmarks}

This section evaluates the existing benchmarks in cybersecurity domain. This can be divided into three primary categories according to what they evaluate: Knowledge and theoretical understanding~\cite{tihanyi2024cybermetric,li2023seceval,bhusal2024secure,simoni2025morse,liu2023secqa}, Cyber Threat Intelligence (CTI) Analysis~\cite{alam2024ctibench,alam2025athenabench,ji2024sevenllm}, Practical Offensive \& Defensive Tasks~\cite{sanz2025cybersecurity,bhatt2023purple,bhatt2024cyberseceval,wan2024cyberseceval,lin2025ircopilot}. An overview of these benchmarks is illustrated in Table \ref{tab:llm_benchmarks}.

\subsection{Knowledge and theoretical understanding}The benchmarks under this taxonomy, primarily use multiple-choice questions (MCQs) or question-answering (QA) formats to evaluate an LLM's theoretical understanding of cybersecurity concepts and frameworks. This includes CyberMetric~\cite{tihanyi2024cybermetric}, which evaluates broad cybersecurity expertise across nine distinct domains: disaster recovery and business continuity planning (BCP), identity and access management (IAM), IoT security, cryptography, wireless and network security, cloud security, penetration testing, and compliance auditing. It is comprised of a multiple-choice question benchmark comprising over 10,000 question–answer pairs. SecEval~\cite{li2023seceval} focuses on theoretical security tasks across 2,000 MCQ instances. Similarly, SecQA~\cite{liu2023secqa} comprises of questions based on content extracted from the textbook ``Computer Systems Security: Planning for Success" (Tolboom 2023). Specialized frameworks like SECURE (Security Extraction, Understanding \& Reasoning Evaluation)~\cite{bhusal2024secure} focuses on the Industrial Control System (ICS) sector. It includes six datasets, including ATT\&CK and CWE extraction tasks (MAET, CWET) to evaluate three types of knowledge evaluation: extraction, understanding, and reasoning. They tested seven LLMs - three commercial and four open source. The study concluded that closed-source models especially ChatGPT-4 and Gemini-Pro, generally exhibit superior performance across most tasks as compared to open source models. Additionally, the MoRSE~\cite{simoni2025morse} is a chatbot architecture that focuses on mitigating hallucination risks in cybersecurity question answering. It uses 600 expert-validated questions and combines RAGAS metrics, GPT-4-based Elo rankings, and LLM-as-judge scoring to assess answer correctness and relevance. It is very evident that these benchmarks are focused on a wide domain of cybersecurity focusing on the in-depth knowledge acquired by LLMs, a not specifically on attack handling capability. The study by Maria Sanz‑Gomez et al.~\cite{sanz2025cybersecurity} reveals that models score high on theoretical knowledge (70–89\% success) but show substantial degradation in complex, multi-step adversarial scenarios (20–40\% success) and worse in robotic targets (22\% success).

\subsection{Cyber Threat Intelligence (CTI) Analysis} Cyber Threat Intelligence (CTI) benchmarks evaluate the practical utility of models in transforming unstructured threat data into actionable security insights. CTIBench~\cite{alam2024ctibench} evaluates LLM performance on cyber threat intelligence (CTI) applications, including mapping CVE descriptions to CWE categories with remediation guidance (CTI-RCM), predicting CVSS severity metrics from vulnerability text (CTI-VSP), extracting MITRE ATT\&CK techniques from threat reports (CTI-ATE), and attributing incidents to threat actors or malware families (CTI-TAA), typically using exact-match and F1-score metrics against curated ground truth labels. AthenaBench~\cite{alam2025athenabench} extends CTIBench by introducing a dynamic CTI evaluation pipeline that de-duplicates CVE/CWE datasets and continuously populates tasks from live sources such as MITRE ATT\&CK and the NVD API, while also introducing a novel task for proposing Risk Mitigation Strategies (RMS). The Objective of the RMS task is to evaluate whether Large Language Models (LLMs) can recommend appropriate MITRE ATT\&CK mitigation strategies based on a description of an observed attack scenario. But here, the attack scenarios are mostly GPT-generated with critical human-in-the-loop stages, rich in technicality and hence cannot be efficient from a first responder perspective. Similarly, in SEVENLLM~\cite{ji2024sevenllm}, raw cybersecurity reports were gathered from vendor websites, and news and select-Instruct method was used wherein GPT-4 generated the specific test questions and their correct answers.

\subsection{Practical offensive and defensive benchmarks} This section marks a transition from static evaluation to execution-based environments where autonomous agents must interact with live systems to solve complex security challenges. CAIBench (Cybersecurity AI Benchmark)~\cite{sanz2025cybersecurity} is a modular meta-benchmark designed to evaluate Large Language Models (LLMs) and AI agents on their labor-relevant capabilities in cybersecurity, addressing the gap between theoretical knowledge and practical execution. It integrates five major evaluation categories—Jeopardy-style CTFs, Attack and Defense CTFs, Cyber Range exercises, knowledge benchmarks, and a novel privacy assessment (CyberPII-Bench)—to systematically measure performance in offensive, defensive, and privacy-preserving operations. From an offensive perspective,the CYBERSECEVAL~\cite{bhatt2023purple,bhatt2024cyberseceval,wan2024cyberseceval} framework measures a model's compliance and helpfulness when prompted to assist in various stages of a cyberattack. Conversely, defensive capabilities are assessed through specialized environments like IRCOPILOT’s IRBench~\cite{lin2025ircopilot}, which utilizes an incremental approach to evaluate the full incident response lifecycle-encompassing detection, containment, and recovery—across 130 sub-tasks on diverse case machines across 27 categories. IRBench could be considered a better benchmark for LLMs for incident response generation. By breaking down the overall response into a series of required sub-tasks like clue acquisition, information analysis, and incident handling, IRBench can quantify both the quality (sub-task completion rate) and the efficiency of the LLM's suggested actions. But this too poses a major challenge. For IRBench to work, you need either an automated system like IRCOPILOT~\cite{lin2025ircopilot} or a Human-in-the-Loop Strategy to execute the model's instructions and feed the results back. While they excel at measuring isolated technical skills or high-level reasoning, they often fail to address the integration challenges of context loss and hallucination in real-time, require complex prerequisites, potentially leading to inaccurate or non-specific recommendations that could exacerbate an active intrusion.

\begin{table*}[t]
\centering
\caption{Comparison of exixsting LLM Benchmarks for Incident Response}
\label{tab:llm_benchmarks}
\small 
\renewcommand{\arraystretch}{1.5} 
\begin{tabularx}{\textwidth}{@{} >{\raggedright\arraybackslash}p{2.2cm} >{\raggedright\arraybackslash}X >{\raggedright\arraybackslash}X >{\raggedright\arraybackslash}X >{\raggedright\arraybackslash}p{2.8cm} @{}}
\toprule
\textbf{Benchmark Name} & \textbf{Assess Legal Correctness?} & \textbf{IR Eval: With Technical Prerequisites (Human/Auto)} & \textbf{IR Eval: No Technical Prerequisites} & \textbf{Intended User} \\
\midrule

AthenaBench~\cite{alam2025athenabench} & 
Partially (evaluates regulatory frameworks like GDPR) & 
Partially (evaluates ``Risk Mitigation Strategy'' but as a static reasoning task) & 
No (Static benchmark focused on reasoning, not on live execution) & 
Security analysts, CTI professionals (3--5 years exp) \\ \addlinespace

CAIBench~\cite{sanz2025cybersecurity} & 
Partially (CyberPII-Bench assesses GDPR compliance) & 
Yes (Interactive A\&D CTFs and Cyber Range exercises via Docker) & 
No (Requires docker infrastructure for execution-based tasks) & 
security researchers, AI agents, penetration testers \\ \addlinespace

CTIBench~\cite{alam2024ctibench} & 
Partially (includes questions on regulations like GDPR) & 
Partially (aims to ``accelerate incident response'' via automated triage) & 
No (Static dataset focused on reasoning and comprehension) & 
CTI practitioners, Security analysts \\ \addlinespace

CyberMetric~\cite{tihanyi2024cybermetric} & 
Partially (includes a domain for ``Compliance/Audit'') & 
No (strictly a multiple-choice Q\&A dataset on general knowledge) & 
No (does not evaluate IR actions) & 
LLM researchers, cybersecurity students/experts \\ \addlinespace

CYBERSECEVAL\\
~\cite{bhatt2023purple,bhatt2024cyberseceval,wan2024cyberseceval} & 
No (focused on secure coding and attack compliance) & 
No (focused on ``Propensity to generate insecure code'' and ``Attack Helpfulness'') & 
No (automated static analysis of code outputs only) & 
LLM designers, software developers \\ \addlinespace

SecQA~\cite{liu2023secqa}  & 
No (focused on computer security principles from a textbook) & 
No (multiple-choice question answering only) & 
No (does not include operational IR tasks) & 
Students, researchers evaluating security understanding \\ \addlinespace

SECURE~\cite{bhusal2024secure} & 
No (focused on technical ICS security and extraction) & 
Partially (evaluates ``Risk Evaluation Reasoning'' for advisories) & 
No (static evaluation of knowledge extraction and reasoning) & 
ICS security analysts, cybersecurity advisors \\ \addlinespace

SEVENLLM~\cite{ji2024sevenllm} & 
No (focused on information extraction from threat reports) & 
Partially (evaluates ``Incident Response Planning'' as a generation task) & 
No (instruction-following evaluation, no live environment) & 
CTI analysts, security event responders \\ \addlinespace

IRBench / IRCOPILOT~\cite{lin2025ircopilot} & 
Partially (includes ``security compliance auditing'' tasks) & 
Yes (Human-in-the-loop executor on TryHackMe/XuanJi platforms) & 
Partially (assesses if LLMs can generate accurate, zero-shot commands) & 
Incident response teams, automated security systems \\ \addlinespace

MoRSE~\cite{simoni2025morse} & 
No (focused on knowledge retrieval for Q\&A) & 
No (chatbot framework for querying security information) & 
No (does not perform IR actions, only provides answers) & 
Security professionals, practitioners seeking technical info \\
\bottomrule
\end{tabularx}
\end{table*}

\section{Conclusion on Existing Benchmarks}
\label{sec:conclusion_benchmarks}
Although we have many existing benchmarks for testing LLMs in the cybersecurity domain, we noticed that they are insufficient to assess the performance of LLMs in supporting law enforcement officers during the response to cybercrime incidents. The limitations can be summarized into four critical gaps, with the most significant being (1) the usability gap for non-specialist first responders. Existing frameworks primarily evaluate prompts that are rich in technicality, often generated by gpt models or gemini. However, they fail to measure whether an LLM can translate these insights into actionable, jargon-free instructions suitable for an officer at a crime scene. Furthermore, (2) the consequence-based risk of hallucinations is not adequately captured by current static metrics. While general benchmarks use F1 or accuracy scores, they do not account for the nature of police operations, where a single hallucinated command, such as an incorrect partition identifier or an invasive memory dump, could lead to the  irreversible destruction of digital evidence. (3) Lack of alignment with procedural frameworks is another issue that happens as benchmarks focus on technical taxonomies like MITRE ATT\&CK rather than the specific Standard Operating Procedures (SOPs). Finally, (4) most existing methodologies pay little attention to legal admissibility, rarely checking whether the technical steps suggested by an LLM can actually be reproduced and presented as valid evidence in court. From these points, we can infer that a specialized evaluation paradigm is necessary to prioritize operational safety and clarity over raw technical throughput.

\section{Potential Mitigation Strategies for the major drawbacks of using LLMs:}
\label{sec:mitigation}

A key challenge in evaluating large language models (LLMs) is randomness inconsistency, wherein repeated executions of the same prompt yield different outputs. Several methodological approaches have been proposed to mitigate this issue and improve robustness. Chain-of-Thought (CoT) prompting encourages structured reasoning by instructing the model to reason step by step before producing a final answer in a specified format~\cite{wang2024mmlu}. Prompt-reverse inconsistency can also be mitigated through CoT prompting. Prompt-reverse inconsistency occurs when an LLM produces conflicting judgments under direct and reverse prompting conditions for the same answer candidates. Negation-based explanatory prompts that clarify the logical structure of reversed queries can be used to address this~\cite{ahn2025prompt}. In addition to this, to make it more reliable, self-consistent sampling generates multiple diverse reasoning paths instead of using a single "greedy" decoding trajectory, and the final answer is determined using majority voting among the sampled outputs~\cite{wang2022self}.

Another strategy is to paraphrase the input query. By applying greedy decoding to each of the model's multiple semantically equivalent reformulations of a given topic, this technique generates reasoning paths. A majority voting method that reduces sensitivity to surface-level linguistic variation is then used to determine the most consistent final response~\cite{chen2024self}. Building on this idea, ranked voting-based self-consistency methods ask the model to produce a ranked list of possible answers rather than just one prediction. The final answer is determined by aggregation methods such as mean reciprocal rank voting, positional voting rules with point allocations, and removing low-vote candidates~\cite{wang2025ranked}. Whereas, the work in \cite{hammar2025incident}, employs an online lookahead planning strategy used for reducing hallucination. In this method, multiple candidate actions are generated, and their future impacts are simulated and the one with the least recovery time is chosen.

Latent Adversarial Paraphrasing (LAP) addresses robustness from a training perspective. It is motivated by the finding that worst-case paraphrases often correlate with high Euclidean distances in the embedding space of the model. LAP uses a learnable latent continuous perturbation inside hidden layers to maximize this distance while preserving semantic coherence via a Lagrangian constraint. Robustness against performance degradation under language variance can be improved by training the model on these adversarial latent paraphrases~\cite{fu2025same}.

\textit{``LLMs struggle with linguistic variability, raising concerns about their generalization abilities and evaluation methodologies. Furthermore, the observed performance drop challenges the reliability of benchmark-based evaluations, indicating that high benchmark scores may not fully capture a model's robustness to real-world input variations~\cite{lunardi2025robustness}''.}

\

\section{conclusion and future scope}
\label{sec:conclusion_and _scope}
The developing decision-support systems for early-stage cybercrime response have been methodically studied in this paper. We have identified a crucial ``usability-infrastructure" gap by assessing the shift from static playbook-driven frameworks to agentic AI architectures. Even though agentic systems provide a high degree of technical autonomy, frontline law enforcement working in environments with inconsistent logging practices and low technical maturity cannot use them.

Our comparative analysis reveals that RAG-based LLM systems currently offer a better pragmatic balance of contextual adaptability and lower deployment barriers. However, as demonstrated in our consistency testing in Section \ref{sec:llm_reliability}, LLMs create significant risks of hallucination and procedural inconsistency. In crucial law enforcement scenarios, an overconfident yet incorrect recommendation can lead to the irreversible loss of volatile evidence or the violation of legal chains of custody.

Moreover, the survey of existing benchmarks further indicates a significant research void: current evaluation frameworks prioritize theoretical knowledge or expert-level technical execution over the clarity, safety, and legal soundness required by a first responder. Consequently, this points out to a need for a specialized evaluation paradigm that measures an AI’s ability to interpret ambiguous, non-technical queries and provide structured guidance.

Future research must focus on the development of robustness-centric benchmarks that address the gaps discussed in Section \ref{sec:conclusion_benchmarks} and explore the true potential of RAG-based LLMs to solve the incident response problem. In parallel, there is significant scope for designing a domain-specific decision-support system that integrates RAG with large language models, grounded in curated databases of real-world cybercrime cases, standard operating procedures, and jurisdiction-specific legal frameworks. Additionally, incorporating explainability and traceability mechanisms will be essential to ensure that every recommendation can be audited and justified in court-admissible contexts. To ensure that the critical first hour of an investigation leads to justice rather than digital attrition, it is necessary to bridge the gap between advanced AI reasoning and the practical, legal, and cognitive constraints of frontline investigation.
\bibliographystyle{IEEEtran}
\bibliography{references}
\end{document}